# Cycling Li-$O_2$ Batteries via LiOH Formation and Decomposition

Tao Liu,[1] Michal Leskes,[1]† Wanjing Yu,[1,2]‡ Amy J. Moore,[1] Lina Zhou,[1] Paul M. Bayley,[1] Gunwoo Kim,[1,2] Clare P. Grey[1]*

**Affiliations:**

[1]Chemistry Department, University of Cambridge, Lensfield Road, CB2 1EW, UK.

[2]Cambridge Graphene Centre, University of Cambridge, Cambridge CB3 0FA, UK.

*Correspondence to: cpg27@cam.ac.uk

†Current address: Weizmann Institute of Science, Department of Materials and Interfaces, Rehovot 76100, Israel.

‡Current address: School of Metallurgy and Environment, Central South University, Changsha 410083, PR China.

**Abstract**: The rechargeable aprotic Li-air ($O_2$) battery is a promising potential technology for next generation energy storage, but its practical realization still faces many challenges. In contrast to the standard Li-$O_2$ cells, which cycle via the formation of $Li_2O_2$, we use a reduced graphene oxide electrode, the additive LiI, and the solvent dimethoxyethane to reversibly form/remove crystalline LiOH with particle sizes > 15 μm during discharge/charge. This leads to high specific capacities, excellent energy efficiency (93.2%) with a voltage gap of only 0.2 V, and impressive rechargeability. The cells tolerate high concentrations of water, water being the dominant proton source for the LiOH; together with LiI it has a decisive impact on the chemical nature of the discharge product and battery performance.

**One-sentence summary**: An efficient aprotic Li-$O_2$ battery cycling via the reversible formation and decomposition of LiOH has been demonstrated in this work.

Rechargeable non-aqueous Li-air ($O_2$) batteries have attracted considerable interest over the past decade, because of their much higher theoretical specific energy than conventional Li ion batteries (1-3). A typical Li-air cell is comprised of a Li metal negative electrode, a non-aqueous $Li^+$ electrolyte and a porous positive electrode. During discharge, $O_2$ is reduced and combines with $Li^+$ at the positive electrode, forming insoluble discharge products (typically $Li_2O_2$) that fill up the porous electrode (4-6). The porous electrode is not the active material, but rather a conductive, stable framework that hosts the reaction products, lighter electrode materials providing higher specific energies. During charge, the previously formed discharge products must be thoroughly removed to prevent the cell from suffocating after a few discharge-charge cycles, the electrode pores becoming rapidly clogged with discharge products and products from unwanted side reactions (7-13).

Several fundamental challenges still limit the practical realization of Li-air batteries (1-3). The first one concerns the reversible capacity (and thus energy density). This is determined by the pore volume of the porous electrode, which limits both the total quantity of the discharge products and how large the discharge product crystals can grow. The ultimate capacity – currently far from being reached – is, in theory, achieved in the extreme case where large single crystals of the discharge product grow to occupy the full geometric volume of the positive electrode. The commonly used mesoporous Super P (SP)/Ketjen carbon electrodes have relatively small pore sizes and volumes, with their crystalline discharge products typically less than 2 μm in size (4-5, 14); this limits the capacity to < 5000 mAh/$g_c$ (typically <1.5 mAh based on 1 mg of carbon and binder) (5, 7-10, 12-13). In addition, uses of smaller pores tend to lead to pore clogging, hindering the diffusion of $O_2$ and $Li^+$ and causing high overpotentials during cycling. Second, severe side reactions can occur on cycling, involving the electrode materials, electrolyte, and intermediate as well as final discharge products (7-13). Major causes of these decomposition reactions include the superoxide ion that forms as an intermediate on reduction of oxygen, which readily attacks most electrolytes (7-9, 15), and the large overpotential on charge, often required to remove the insulating discharge products, which results in oxidation of cell components such as the host electrode (10-13). Previous studies (10-13) suggest that 3.5 V (vs. Li/$Li^+$) represents the maximum voltage that carbon-based electrodes can tolerate without significant side reactions. Third, the large hysteresis seen between charge and discharge (up to 2 V) (4-13), results in extremely low energy efficiencies, limiting the use of this battery in practical applications. Finally, the cells are very sensitive to moisture and carbon dioxide (16-19): the more stable LiOH and Li carbonate phases are formed, which gradually accumulate in the cell, resulting in battery failure. Moisture and $CO_2$ also have deleterious effects on the Li-metal anode (1-3).

A number of strategies have been proposed to reduce the voltage hysteresis, involving the use of electrocatalysts (20-27), porous electrode structures (28-30) and redox mediators (31-35). Notably, soluble redox mediators, such as tetrathiafuvalene (TTF) (32) and LiI (34), have been used to reduce the overpotential of the charge process, the overall voltage hysteresis dropping to around 0.5 V (34). Their operation relies on the electrochemical oxidation of the mediator which itself then chemically decomposes the $Li_2O_2$. The charge voltage is thus tuned close to the redox potential of the mediator. For discharge, the ethyl viologen redox couple (31) has also been used to reduce $O_2$ in the liquid electrolyte rather than on the solid electrode surface, again to help prevent rapid blocking of the solid electrode surface by $Li_2O_2$. Here we use the redox mediator LiI and report a lithium-oxygen battery with an extremely high efficiency, large capacity and a very low overpotential. Of note, this battery cycles via LiOH formation, not $Li_2O_2$, and is able to tolerate large quantities of water. This current work addresses directly a number of critical issues associated with this battery technology.

A $Li-O_2$ battery was prepared by using Li metal anode, 0.25 M lithium bis (trifluoromethyl) sulfonylimide (LiTFSI) / dimethoxyethane (DME) electrolyte with 0.05 M LiI additive, and a variety of different electrode structures (**S1**). Hierarchically macroporous reduced graphene oxide (rGO) electrodes (binder-free) are used because they are light, conductive and have a large pore volume that can potentially lead to large capacities. Mesoporous SP carbon and mesoporous titanium carbide (TiC) (36) electrodes are studied for comparison. Cyclic voltammetry (CV) measurements confirmed that rGO, SP and TiC electrodes all exhibit good electrochemical stability within a voltage window of 2.4-3.5 V in a LiTFSI/DME electrolyte and can be used to reversibly cycle LiI ($I_3^- + 2e^- \leftrightarrow 3I^-$) (37) (**S2**).

In the absence of LiI, cells using either mesoporous TiC or macroporous rGO showed much smaller overpotentials during charge, in comparison to that obtained with the SP electrode (Figure 1A); these decreases in overpotential are tentatively ascribed to the higher electrocatalytic activity of TiC (38) and the faster diffusion of $Li^+$ and solvated $O_2$ within the micron-sized pores of the rGO electrodes (see supplemental information, Figure **S1**). Addition of LiI to the SP electrode led to a noticeable drop in the overpotential over that seen with SP only, suggesting that the polarization during charge is largely caused by the insulating nature of the discharge products. The charge voltage profile is not, however, flat, but gradually increases as the charge proceeds to above 3.5 V. By contrast, when LiI is used with hierarchically macroporous rGO electrodes, a remarkably flat process is observed at 2.95 V, representing a further reduction in overpotential by ~0.5 V over that seen for SP. This reduction is ascribed, at least in part, to the interconnecting macroporous network of rGO, which allows for much more efficient mediator diffusion than in the mesoporous SP electrode, even when the macropores are filled with insoluble discharged products.

The observation that the LiI/DME Li-$O_2$ cell charges at 2.95 V is of note, as it is slightly below the thermodynamic voltage of 2.96 V of the Li-$O_2$ reaction. During charge, the redox mediator is thought to be first electrochemically oxidized on the electrode (32), this oxidized form then helping to chemically decomposes the discharge product. The charge voltage then reflects the redox potential (vs. Li/$Li^+$) of the $I^-/I_3^-$ redox mediator in the electrode/electrolyte system rather than the redox potential associated with the oxidation of the solid discharge product. A low redox potential of a mediator is important for the long-term stability of the Li-$O_2$ cell.

To investigate factors affecting the redox potential, LiI was cycled galvanostatically in an Ar atmosphere with different electrode/electrolyte combinations (Fig. 1B). The electrolyte solvent has a larger effect on the redox potential of the $I^-/I_3^-$ couple than the electrode material, with the DME electrolyte consistently exhibiting lower charge voltages than TEGDME (tetraethylene glycol dimethyl ether) for all three electrodes. In addition, the voltage gaps between the charge and discharge plateaus are smaller for DME than TEGDME electrolytes, consistent with the smaller voltage separations seen between the redox peaks in their respective CV curves (**S2**). The discharge capacity is always smaller than the previous charge capacity for all cells (Fig. 1B), indicating that some mediators after being oxidized have diffused into the bulk electrolyte. This observation is more prominent with DME, suggesting faster mediator diffusion in DME than in TEGDME.

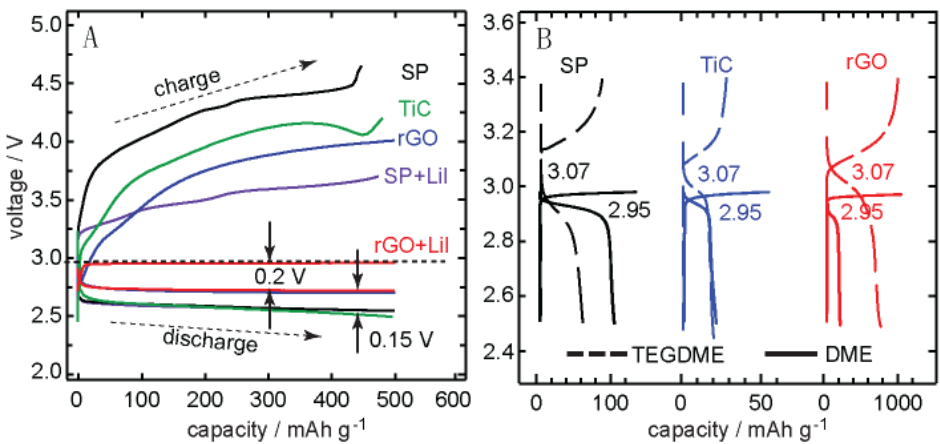

Fig. 1 (A) Discharge-charge curves for Li-O$_2$ cells using mesoporous SP and TiC, and macroporous rGO electrodes, with capacities limited to 500 mAh/g (based on the mass of carbon or TiC) and 0.25 M LiTFSI/DME electrolyte. For SP and rGO electrodes, 0.05 M LiI was added to the LiTFSI/DME electrolyte in a second set of electrodes (purple and red curves). All cells in (A) were cycled at 0.02 mA/cm$^2$. The horizontal dashed line represents the position (2.96 V) of the thermodynamic voltage of a Li-O$_2$ cell. (B) Galvanostatic charge-discharge curves of cells containing 0.05 M LiI and 0.25 M LiTFSI, cycled under an Ar atmosphere with different electrode/electrolyte solvent combinations with a current of 0.2 mA/cm$^2$. The crossing points (appropriate voltages labeled) of the charge-discharge curves indicate the positions of the redox potential of I$^-$/I$_3^-$ in the specific electrode-electrolyte system. A direct comparison of capacities between LiI in Ar and Li-O$_2$ cells is given in **S3**.

The discharge overpotential for rGO-based Li-oxygen cells also decreases by 0.15 V (marked by arrows in Fig. 1A), from 2.6 (SP/TiC) to 2.75 V (rGO) regardless of the use of LiI. Overall, the voltage gap becomes only 0.2 V (indicated by arrows), representing an ultrahigh energy efficiency of *93.2%*.

Surprisingly, XRD patterns (Fig. 2A) for the rGO electrodes cycled with LiI show that LiOH is the only observed crystalline discharge product; LiOH is then removed after a full charge. This is confirmed in the ssNMR measurements (Fig. 2B), where a single resonance due to LiOH is observed at -1.5 ppm and at 1.0 ppm in the $^1$H and $^7$Li MAS ssNMR spectra (13, 39), respectively (further corroborated by the $^7$Li static NMR spectrum in **S4**). After charge, the $^1$H and $^7$Li LiOH resonances are no longer visible. We emphasize that without added LiI, the predominant discharge product for rGO electrodes is Li$_2$O$_2$ (**S5**), the chemistry radically changing when 0.05 M LiI is added to the DME electrolyte.

Figure 2 (C-D) shows optical and SEM images of electrodes during the 1st cycle. After discharge, the electrode surface is completely covered by LiOH agglomerates, tens of microns in size, and the color of the electrode has changed from black to white. When the interior of the electrode was investigated, many crystalline "flower-like" agglomerated LiOH particles were observed within the graphene macropores. Although these particles are more than 15 μm in diameter (**S6**), much bigger than the $Li_2O_2$ toroids (**S5**), they are in fact formed from thin sheet primary building blocks, resulting in a more open (porous) structure. The large LiOH agglomerates efficiently fill up the pore volume available in the hierarchical macroporous electrode, leading to much larger capacities (**S6**). When TEGDME was used as the electrolyte solvent, the discharge product, although still LiOH, now forms a thin film on the rGO electrode surface (**S7**). After charge in DME, the hierarchically macroporous structure reappeared and the electrode turned black again (Fig. 2(C)). Higher magnification SEM images revealed very small traces of residual LiOH on the electrode surface (**S8**). We found that the reversible formation and removal of LiOH with the LiI mediator is not restricted to rGO electrodes, mesoporous SP electrodes showing similar results (**S9**) but with larger overpotentials and lower capacities.

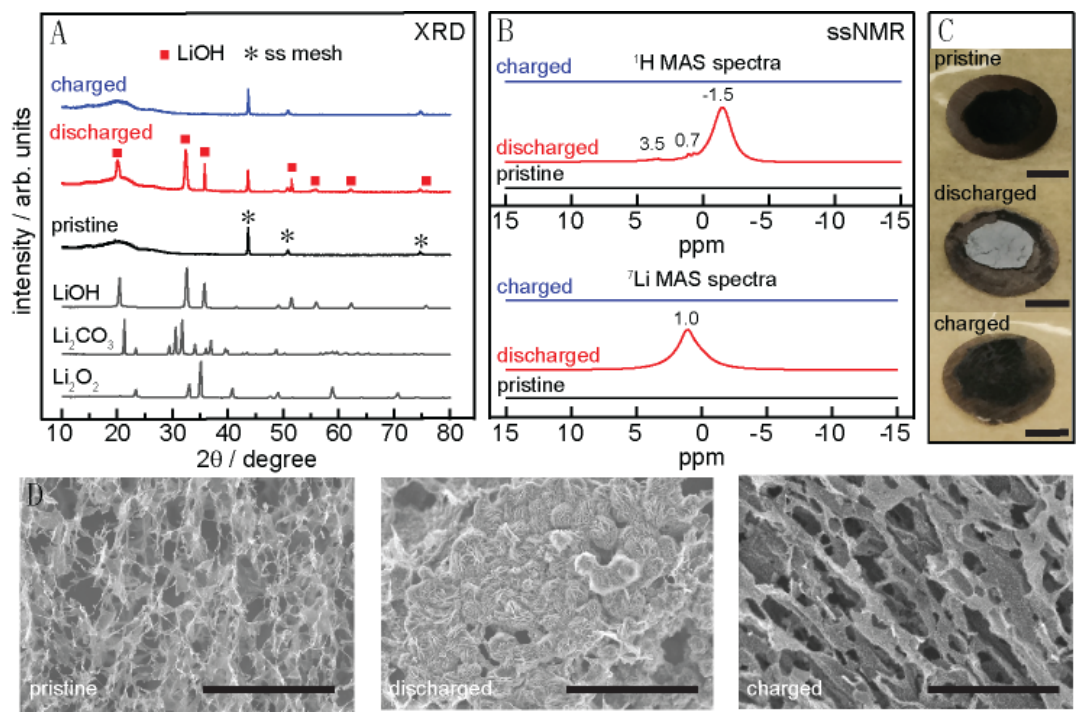

Fig. 2 XRD patterns (A) and $^1H$ and $^7Li$ ssNMR spectra (B) comparing a pristine rGO electrode to electrodes at the end of discharge and charge in a 0.05 M LiI/0.25 M LiTFSI/DME electrolyte. (Electrochemistry of the cells is in Figure S10). The spectra are scaled according to the mass of the pristine electrode and number of scans. Asterisks in (A) represent diffraction peaks from the stainless steel mesh. $^1$H resonances of proton-containing functional groups in the pristine rGO electrode are not visible in the $^1$H ssNMR spectrum in (B) since they are very weak in comparison to the LiOH signal. The weaker signals at 3.5 and 0.7 ppm are due to DME and grease/background impurity signals, respectively. Optical (C) and SEM images (D) of pristine, fully discharged and charged rGO electrodes obtained with a 0.05 M LiI/0.25 M LiTFSI/DME electrolyte in the 1$^{st}$ cycle. The scale bars are all 5 mm and 20 μm in the optical and SEM images, respectively.

Kang and coworkers (34) previously reported a highly rechargeable Li-O$_2$ cell using carbon nanotubes and the mediator LiI (0.05 M) in a TEGDME-based electrolyte, ascribing the electrochemistry to the formation and decomposition of Li$_2$O$_2$. Sun *et al.* (35) recently pointed out, however, that LiOH, rather than Li$_2$O$_2$, is the dominant discharge product when 0.05 M LiI mediator was added to the TEGDME electrolyte; LiOH was still present in the Super P electrode used in their study after charge and they suggested that LiOH could not be decomposed by the mediator. In our work, we have seen clear evidence that the discharge product is overwhelmingly LiOH and importantly it can be removed at low potentials of around 3 V.

In a redox mediated Li-O$_2$ system, the effective removal of the insulating discharge products is affected by a few factors: (1) availability of bare electrode surfaces to oxidize the mediator during charge; (2) whether the discharge product is uniformly distributed throughout the electrode; (3) efficient diffusion of the oxidized mediator from electrode surfaces (that supply/remove electrons) to the discharge product. TEGDME, being a more viscous solvent than DME, will lead to more sluggish I$_3^-$ and O$_2$ diffusion. When it is used with mesoporous (rather than macroporous) electrodes, the discharge product tends to concentrate on the electrode surface facing the gaseous O$_2$ reservoir, with its concentration dropping noticeably in the electrode interior (the reaction zone problem for Li-air batteries (40)). The much more soluble LiI, however, is likely to be uniformly oxidized across the whole thickness of the electrode during charge. Consequently, for equal capacities for discharge and charge, the oxidized mediator (I$_3^-$) formed during charge may remain in excess in the electrode interior regions where the discharge product is scarce; similarly, discharge product may be left unreacted at regions close to the O$_2$/electrolyte interface where the discharge product is abundant. The remaining mediator in the oxidized form (I$_3^-$) will then be reduced during the next discharge, resulting in a voltage plateau at its redox potential in addition to that due to oxygen reduction. This unbalanced distribution of the mediator LiI and the discharge product LiOH across the thickness of the mesoporous electrode may be a cause of the unreacted LiOH and iodine-dominated electrochemistry observed in the work by Sun and coworkers (34). Furthermore, the thin film morphology of the discharge product formed in TEGDME-based electrolyte (Fig. S7 and ref (35)), effectively passivates the electrode surface. As a result, triiodide anions may first form on bare electrode surfaces that are distant from the discharge products. They then need to diffuse to the passivated regions to remove LiOH reducing the efficiency of LiOH removal, providing another explanation for the observed residual LiOH. Here we used a macroporous rGO electrode and DME, which provides faster mediator and $O_2$ diffusion, and higher $LiO_2$ solubility leading to a more uniform $Li-O_2$ reaction during discharge and larger reversible capacities.

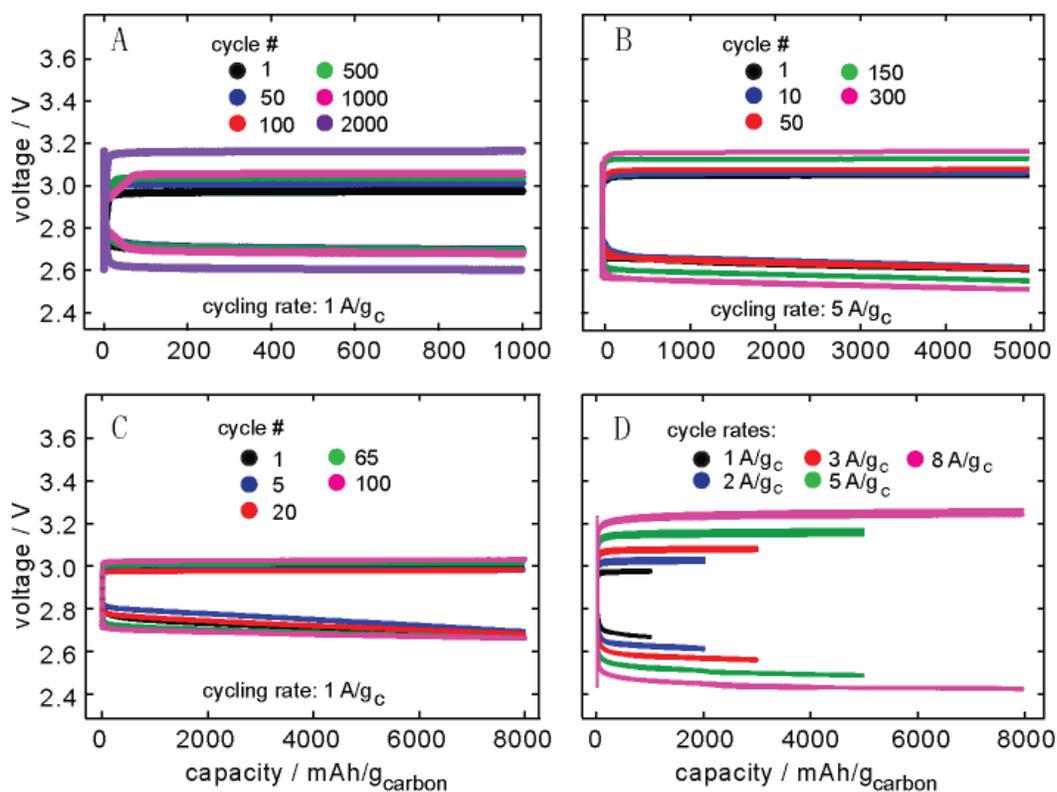

Fig. 3 Discharge-charge curves for $Li-O_2$ batteries using rGO electrodes and 0.05 M LiI/0.25 M LiTFSI/DME electrolyte with capacity limits of 1000 mAh/$g_c$ (A), 5000 mAh/$g_c$ (B), and 8000 mAh/$g_c$ (C), as a function of rate (D); 3 cycles were performed for each rate in (D). The cell cycle rate is based on the mass of rGO, i.e., 5 A/$g_c$ is equivalent to 0.1 mA/cm$^2$.

Figure 3 shows the electrochemical performance of the $Li-O_2$ battery. When limiting the specific capacity to 1000, 5000 and 8000 mAh/$g_c$, the cells show *no* capacity fade with little increase in voltage polarization after 2000, 300 and 100 cycles, respectively (Fig. 3A-C). Higher capacities >

20,000 mAh/$g_c$ have also been demonstrated (**S5** and **S11**). When cycled at 1 A/$g_c$ (Fig. 3A and C), the voltage gap is only ~0.2 V; at higher rates the gaps widen (Fig. 3D), increasing to 0.7 V at 8 A/$g_c$. Furthermore, at this higher rate, the cell is polarized each cycle (**S11**) and after 40 cycles the electrode surface is covered by a large number of particles (with morphologies unlike those of LiOH observed at lower currents), which do not seem to be readily removed during charge at these voltages. At these higher overpotentials, more substantial parasitic reactions likely occur, rapidly polarizing the cell by increasing its resistance and impeding the electron transfer across the electrode-electrolyte interface. A narrower operating electrochemical window within 2.96±0.5 V is key for the prolonged stability of rGO electrodes.

The sensitivity of the cell to water was explored by either deliberately adding ~45,000 ppm of $H_2O$ (37 mg per 783 mg of DME) to the electrolyte or cycling cells under a humid $O_2$ atmosphere. In both cases, no appreciable change in the electrochemical profile was observed (**S12**), compared to that using a nominally dry electrolyte. Furthermore, the added water is found to promote growth of even larger LiOH crystals > 30 μm (**S13**).

Although a certain level of scattering in the total capacity is observed, probably due to variations in the electrode structure, the cell capacity is typically within 25,000-40,000 mAh/$g_c$ (i.e. 2.5-4.0 mAh) range for an rGO electrode of 0.1 mg and 200 μm thick. After discharge, the weight of an electrode removed from the stainless steel mesh was about 1.5 mg (2.7 V, 3.2 mAh), giving a specific energy of 5760 Wh/kg (see Section 13 of the SI).

The mechanism of $O_2$ reduction in aprotic Li-$O_2$ batteries has been extensively discussed. Many authors (5, 41-42) have shown that the ability of an electrolyte to solvate $O_2^-$ (characterized by the Guttman Donor Number, DN) is important in governing the discharge reaction mechanism. Higher $LiO_2$ solubility favors a solution precipitation mechanism leading to large toroidal $Li_2O_2$ crystals and thus higher discharge capacity; lower $LiO_2$ solubility tends to drive a surface mechanism where $Li_2O_2$ forms thin films on the electrode surface and a lower capacity. Because of its intermediate DN, solution precipitation and surface reduction mechanisms can occur simultaneously in DME (41).

With added LiI, although LiOH rather than $Li_2O_2$ is the prevailing discharge product, many parallel phenomena are observed: the similar discharge voltages (2.75 V, Fig. 1A) observed with and without the added LiI suggests that the first step on discharge is an electrochemical reaction, where $O_2$ is reduced on the electrode surface to form $LiO_2$. It is unlikely that $O_2$ is reduced to $O_2^{2-}$ via two electron electrochemical steps or even dissociatively reduced to $O^{2-}$ (or LiOH) via four electron electrochemical steps. Subsequent conversion of $LiO_2$ to LiOH is proposed to be a chemical process that occurs via a solution mechanism. Strong support for a solution mechanism comes from the observation that LiOH grows on both the electrode and insulating glass fiber separators (**S6**), the latter not being electrically connected to the current collector. This process must involve the iodide redox mediator, because in its absence $Li_2O_2$ is formed, even in cells with high moisture contents.

A key question is the origin of the $H^+$ in the formed LiOH, potential sources being the DME electrolyte, surface functional groups of rGO, and moisture in the cell (**S14**). To investigate this, NMR measurements were conducted on two sets of discharged electrodes from cells that were prepared using either deuterated DME and or deuterated water (**S15-18**). When deuterated DME was used, only a very small quantity of LiOD was detected in the $^2$H NMR spectra (**S15-16, 18**), the dominant product being LiOH. By contrast, when $D_2O$ was added to the protonated DME electrolyte, a significant amount of LiOD was observed (**S17**), LiOH only being a minor signal in the corresponding $^1$H NMR spectrum. In summary, these experiments clearly demonstrate that: (1) although DME can be a potential $H^+$ source for LiOH, it is by no means the dominant one; (2) the added water preferentially supplies $H^+$ to form LiOH, substantially minimize DME decomposition (**S17-18**); (3) even our nominally dry cells contain sufficient water to promote LiOH formation. This latter statement is consistent with the formation of large toroidal $Li_2O_2$ particles in the absence of LiI (**S5**), earlier work showing that this requires water levels of > 500 ppm (42).

$^1$H NMR spectroscopy was used to quantify the number of moles of LiOH formed on discharge (with added LiI mediator). The (molar) ratio of electrons consumed on discharge to LiOH formed was close to 1:1 within the errors of the measurements (**S19**), supporting the proposal that LiOH is formed in stoichiometric quantities (i.e., is not a minor side-product).

During charge, an iodine-mediated LiOH decomposition reaction is observed at ~3 V, a clear distinction of our work from others (34-35). Given that this charge voltage overlaps with that for I$^-$/I$_3^-$ itself (measured with Ar gas in Fig. 1B), the first step should involve the direct electrochemical oxidation of I$^-$ to I$_3^-$. We suggest that the next step involves the chemical oxidation of LiOH by I$_3^-$ to form O$_2$ and H$_2$O. To confirm this hypothesis, a pre-discharged Li-O$_2$ cell (with LiI) was charged in a pure *argon* atmosphere and the gas atmosphere in the cell after charge was monitored by mass spectrometry. A clear O$_2$ signal (**S21**) was detected consistent with the proposal that LiOH decomposition via an O$_2$ evolution reaction. The discharge and charge reactions are schematically represented in Fig. 4. We stress, however, that the equilibria that occur in the presence of oxygen, water and iodine are complex often involving a series of polyanions (including IO$^-$ and its protonated form); further mechanistic studies are required to understand the role of these complex equilibria in the redox processes.

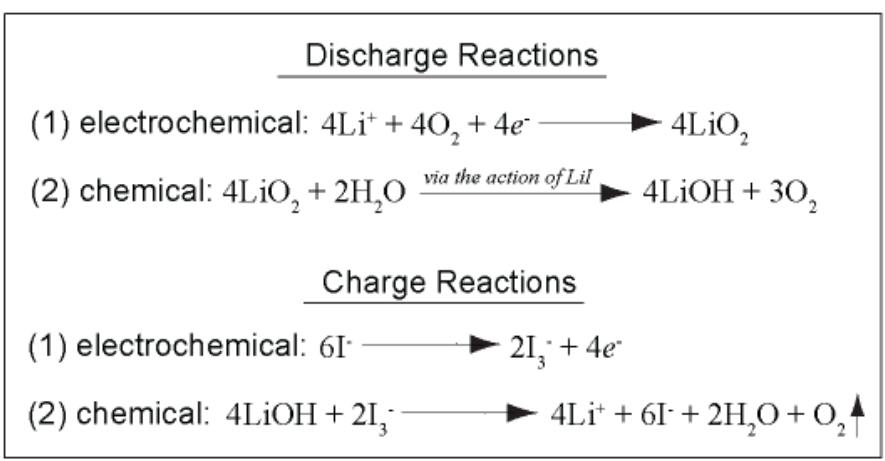

Fig. 4 Schematic mechanisms for the formation and removal of LiOH in iodide redox-mediated Li-O$_2$ cells in the presence of water. The electron/LiOH molar ratios during discharge and charge are both equal to 1.

In conclusion, by using an rGO electrode and the redox mediator LiI, in a DME-based electrolyte, we have demonstrated a highly efficient, rechargeable Li-O$_2$ battery with extremely large capacities. Its operation involves the reversible formation and removal of LiOH crystals. The role of the additive, LiI, is threefold. First it operates as a redox mediator whose redox potential can be tuned by using different electrolyte solvents and electrode structures; this redox potential guides the charge voltage and thus affects the cycling stability of the cell. Second, LiI, together with another additive H$_2$O, impacts the chemical nature and physical morphology of the discharge products, inducing the growth of large LiOH crystals that efficiently take up the pore volume of macroporous rGO electrodes; this is the origin of the observed large capacity. Finally, it enables a chemical pathway to remove LiOH at low overpotentials. Consequently the cell becomes insensitive to relatively high levels of water contamination. The hierarchically macroporous rGO electrode is also an important factor for the high efficiency and capacity. Not only does the rGO framework provide efficient diffusion pathways for all redox active species in the electrolyte and hence, a reduced cell polarization and flatter electrochemical profile, it also permits the growth of LiOH crystals of tens of microns in size, resulting in a capacity that is much closer to the theoretical value of lithium-air batteries. These desirable features were not observed for Li-$O_2$ cells with mesoporous SP electrodes, even when the same electrolyte was used. The combination of electrolyte additives, the porous electrode structure and the electrolyte solvent, synergistically, not only determines the chemical nature of the discharge product, but also governs the physical size and morphology of it, playing a decisive factor in the capacity and rechargeability of the resulting Li-$O_2$ battery. In a broader sense, this work inspires ways to remove other stable, detrimental chemicals, such as $Li_2CO_3$, relevant to cycling Li-air batteries in real practical conditions.

**Acknowledgements**

This work was partially supported by the Assistant Secretary for Energy Efficiency and Renewable Energy, Office of Vehicle Technologies of the U.S. Department of Energy under Contract No. DE-AC02-05CH11231, under the Batteries for Advanced Transportation Technologies (BATT) Program subcontract #7057154 (WY, ML, PB), EPSRC (TL), Johnson Matthey (AM) and Marie Curie Actions (PB and ML). Dr. M.T.L. Casford is thanked for many useful discussions.

**References and Notes:**

1. G. Girishkumar, B. McCloskey, A. C. Luntz, S. Swanson, W. Wilcke, *J. Phys. Chem. Lett.* **1**(14), 2193 (2010).

2. P. G. Bruce, S. A. Freunberger, L. J. Hardwick, J. M. Tarascon, *Nat. Mater.* **11**, 19 (2012).

3. Y. C. Lu, *et al.*, *Energy Environ. Sci.* **6**, 750 (2013).

4. R. R. Mitchell, B. M. Gallant, Y. Shao-Horn, C. V. Thompson, *J. Phys. Chem. Lett.* **4**, 1060 (2013).

5. B. D. Adams, *et al*. *Energy Environ. Sci.* **6**, 1772 (2013).

6. B. M. Gallant, *et al.*, *Energy Environ. Sci.* **6**, 2518 (2013).

7. S. A. Freunberger, *et al.*, *J. Am. Chem. Soc.* **133**, 8040 (2011).

8. B. D. McCloskey, D. S. Bethune, R. M. Shelby, G. Girishkumar, A. C. Luntz, *J. Phys. Chem. Lett.* **2**, 1161 (2011).

9. S. A. Freunberger, *et al.*, *Angew. Chem. Int. Ed.* **50**, 8609 (2011).

10. B. D. McCloskey, *et al.*, *J. Phys. Chem. Lett.* **3**, 997 (2012).

11. B. M. Gallant, *et al.*, *J. Phys.Chem. C* **116**, 20800 (2012).

12. M. M. Ottakam Thotiyl, S. A. Freunberger, Z. Peng, P. G. Bruce, *J. Am. Chem. Soc.* **135**, 494 (2013).

13. M. Leskes, A. J. Moore, G. R. Goward, C. P. Grey, *J. Phys. Chem. C* **117**(51), 26929 (2013).

14. D. Zhai, *et al.*, *J. Am. Chem. Soc.* **135**, 15364 (2013).

15. E. Nasybulin, *et al.*, *J. Phys. Chem. C*, **117**, 2635 (2013).

16. S. R. Gowda, A. Brunet, G. M. Wallraff, B. D. McCloskey, *J. Phys. Chem. Lett.* **4**, 276 (2013).

17. H. K. Lim, *et al.*, *J. Am. Chem. Soc.* **135**, 9733 (2013).

18. Y. Liu, R. Wang, Y. Lyu, H. Li, L. Chen, *Energy Environ. Sci.* **7**, 677 (2014).

19. Z. Guo, X. Dong, S. Yuan, Y. Wang, Y. Xia, *J. Power Sources* **264**, 1 (2014).

20. Y. C. Lu, H. A. Gasteiger, Y. Shao-Horn, *J. Am. Chem. Soc.* **133**, 19048 (2011).

21. B. D. McCloskey, *et al.*, *J. Am. Chem. Soc.* **135**, 18038 (2011).

22. S. H. Oh, L. F. Nazar, *Adv. Energy Mater.* **2**, 903 (2012).

23. S. H. Oh, R. Black, E. Pomerantseva, J. H. Lee, L. F. Nazar, *Nat. Chem.* **4**, 1004 (2012).

24. J. Lu, *et al.*, *Nat. Comm.* **4**, 1 (2013).

25. H. G. Jung, *et al.*, *ACS Nano* **7**, 3532 (2013).

26. E. Yilmaz, C. Yogi, K. Yamanaka, Y. Ohta, H. R. Byon, *Nano Lett.* **13**, 4679 (2013).

27. B. Sun, X. Huang, S. Chen, P. Munroe, G. Wang, *Nano Lett.* **14**, 3145 (2014).

28. J. Xiao, *et al.*, *Nano Lett.* **11**, 5071 (2011).

29. Z. L. Wang, D. Xu, J. J. Xu, L. L. Zhang, X. B. Zhang, *Adv. Funct. Mater.* **22**, 3699 (2012).

30. H. D. Lim, *et al.*, *Adv. Mater.* **25**, 1348 (2013).

31. M. J. Lacey, J. T. Frith, J. R. Owen, *Electrochem. Commun.* **26**, 74 (2013).

32. Y. Chen, S. A. Freunberger, Z. Peng, O. Fontaine, P. G. Bruce, *Nat. Chem.* **5**, 489 (2014).

33. D. Sun, *et al.*, *J. Am. Chem. Soc.* **136**, 8941 (2014).

34. H. D. Lim, *et al.*, *Angew. Chem. Int. Ed.*, **126**, 4007 (2014).

35. W. J. Kwak, *et al.*, *J. Mater. Chem. A*, **3**, 8855 (2015).

36. M. M. Ottakam Thotiyl, *et al.*, *Nat. Mater.* **12**, 1050 (2013).

37. K. J. Hanson, C. W. Tobias, *J. Electrochem. Soc.* **134**, 2204 (1987).

38. B. D. Adams, *et al.*, *ACS Nano*, **8**, 12483 (2014).

39. M. Leskes, *et al.*, *Angew. Chem. Int. Ed.* **51**, 8560 (2012).

40. S. S. Zhang, D. Foster, J. Read, *J. Power Sources,* **195**, 1235 (2010).

41. L. Johnson, *et al.*, *Nat. Chem.* **6**, 1091 (2014).

42. N. B. Aetukuri, *et al.*, *Nat. Chem.* **7**, 50 (2015).

43. W. S. Hummers Jr., R. E. Offeman, *J. Am. Chem. Soc.*, **80** (6), 1339 (1958).

44. Y. Zhao, L. Wang, H. R. Byon, *Nat. Commun.*, **4**, 1896 (2013).

45. L. H. Saw, Y. Ye, A. A. O. Tay, *Energy Convers. Manage.* **75**, 162 (2013).

46. P. Stevens, *et al.*, *ECS Trans.* **28** (32), 1 (2010).

47. R. S. Assary, K. C. Lau, K. Amine, Y. K. Sun, L. A. Curtiss, *J. Phys. Chem. C*, ***117***(16), 8041 (2013).